\documentclass{ws-procs975x65}

\def\cA {\cal A}

\def\be {\begin{equation}}
\def\ee {\end{equation}}
\def\beq{\begin{equation}}
\def\eeq{\end{equation}}
\def\bea {\begin{eqnarray}}
\def\eea {\end{eqnarray}}
\def\br{\begin{eqnarray}}
\def\er{\end{eqnarray}}
\def\bc {\begin{center}}
\def\ec {\end{center}}
\def\bi {\begin{itemize}}
\def\ei {\end{itemize}}
\def\benu{\begin{enumerate}}
\def\eenu{\end{enumerate}}
\newcommand{\bdm}{\begin{displaymath}}
\newcommand{\edm}{\end{displaymath}}

\def\la {\label}
\def\le {\left}
\def\ri {\right}
\def\l{\left}
\def\r{\right}

\def\laq{\hbox{~}\raise 0.4ex\hbox{$<$}\kern -0.8em\lower 0.62ex\hbox{$\sim$}\hbox{~}}
\def\gaq{\hbox{~}\raise 0.4ex\hbox{$>$}\kern -0.7em\lower
 0.62ex\hbox{$\sim$}\hbox{~}}

\newcommand{\SEN}{S_{_{\rm ent}}}

\newcommand{\SBH}{S_{_{\rm BH}}}

\newcommand{\rHo}{r_{_{\rm H}}}

\begin{document}

\title{Entanglement and corrections to Bekenstein-Hawking entropy}

\author{{\bf Saurya Das}$^{(1)}$, {\bf S. Shankaranarayanan}$^{(2,3)}$\footnote{Speaker, E-mail:~shanki@iisertvm.ac.in} and {\bf Sourav Sur}$^{(4)}$}
\address{${}^{(1)}$University of Lethbridge, 4401 University Drive, 
Lethbridge, Alberta, Canada T1K 3M4\\
${}^{(2)}$~Institute of Cosmology and Gravitation, University of Portsmouth, 
Portsmouth, U.K.\\
${}^{(3)}$ School of Physics, Indian Institute of Science Education and 
Research-Trivandrum,\\ CET campus, Thiruvananthapuram 695 016, India.\\
${}^{(4)}$ Theory Division, Saha Institute of Nuclear Physics
Bidhannagar, Kolkata 700064, India}
\begin{abstract}
In this talk, we focus on the corrections to Bekenstein-Hawking 
entropy by associating it with the entanglement between degrees of freedom 
inside and outside the horizon. Using numerical techniques, we 
show that the corrections proportional to fractional power of area 
result when the field is in a superposition of ground and excited states.
We explain this result by identifying that the degrees of freedom
contributing to such corrections are different from those contributing
to Bekenstein-Hawking entropy.
\end{abstract}

\bodymatter

\bigskip

The entropy of a black-hole (BH), given by the well-known
Bekenstein-Hawking relation \cite{Bek} (the so-called {\it area law}
of black-hole physics)
{\small
\be \la{eq:SBH}
S_{_{\rm BH}} = \le(\frac{k_{_{B}}}{4}\ri) 
\frac{\cA_{\rm H}}{\ell_{_{\rm Pl}}^2} ,
\quad \le({\ell_{_{\rm Pl}}} \equiv \sqrt{\frac{G\hbar}{c^3}} = 
\mbox{Planck length},~\cA_{\rm H} = \mbox{Horizon area}\ri) ,
\ee
}
is characteristically distinct from that of other physical systems,
e.g. ideal gases, because it is finite only for a quantum description
of gravity and(or) matter fields.  It is therefore expected that any
viable quantum gravity (QG) model should explain the origin of the BH
entropy, identifying the microscopic degrees of freedom (DOF) that
give rise to it. The area law is shown to hold in several approaches,
starting from those that count microstates assuming fundamental
structures (string, loop, etc.)  \cite{stringsetc}, to the
entanglement of quantum modes inside and outside of the BH horizon
\cite{bomb,sred,sdss}.

However $S_{_{\rm BH}}$ in Eq. (\ref{eq:SBH}) being, by origin, a
semi-classical result, there is no reason to believe it to be the
complete answer conceivable from a correct QG theory, and valid even
for small Planck-sized BHs (i.e., ${\cA_{\rm H}} \sim \ell_{_{\rm
    Pl}}^2$). Therefore, it is imperative for any approach to QG to go
beyond $\SBH$ and provide generic subleading corrections. Another
crucial thing is to identify which of the quantum DOF contribute to
$\SBH$ and which to the corrections, for if these DOF are different
then possibly one may isolate their individual contributions for a
deeper view on the mechanism of entropy generation. This can be
illustrated in the quantum entanglement approach
\cite{sdssdof,sdssssdof,sdssss} of BH entropy, which predicts generic
power-law corrections to $S_{_{\rm BH}}$, as we discuss below.

\vskip 0.05in
\noindent
\underline{\bf Entanglement and its connection to black-hole entropy:}

On a product of two Hilbert spaces ${\cal H} = {\cal H}_1 \otimes
{\cal H}_2$, if a wave-function $\Psi$ cannot be factorized into
wave-functions $\Psi_1$ and $\Psi_2$ on ${\cal H}_1$ and ${\cal H}_2$
respectively, i.e., $\Psi \neq \Psi_1 \otimes \Psi_2$, then the states
described by $\Psi_1$ and $\Psi_2$ are said to be {\it entangled}.
The associated entropy, called the entanglement entropy is given by
the Von Newmann relation $\SEN = - {\rm Tr} [\rho_\alpha
\ln(\rho_\alpha)]$, where $\rho_{\alpha} \, \l(\alpha \in \l\{1,
2\r\}\r) \,$ is the reduced density matrix of a system in any one of
the subspaces ${\cal H}_1$ and ${\cal H}_2$.

$\SEN$, being a quantum effect without classical analogue and
associated with the existence of a horizon, similar to BH entropy, may
serve as the source of the latter \cite{sdssss}.

\vskip 0.05in
\noindent
\underline{\bf Entanglement entropy computation --- setup and assumption:}

We consider the propagation of a massless scalar
field\footnote{Motivated from the point of view of the metric
  perturbations of black-hole space-times\cite{chandra}, that
  correspond to test scalar fields in these space-times
  \cite{sdssss}.}  $(\varphi)$ in an asymptotically flat BH
background, and choose to work in the Lema\^itre coordinates
$(\tau,\xi,\theta,\phi)$ rather than in the Schwarzschild coordinates
$(t,r,\theta,\phi)$ because of some advantages: (i) the BH line
element in Lema\^itre coordinates is non-singular at the horizon
($\rHo$), unlike that in Schwarzschild coordinates, (ii) $\xi$ (or
$\tau$) is space(or, time)-like everywhere, as opposed to $r$ which is
space-like only for $r > \rHo$. However, the scalar field Hamiltonian
in Lema\^itre coordinates is explicitly time-dependent.

We assume that the Hamiltonian evolves adiabatically,
implying that the evolution of the late-time modes leading
to Hawking particles are negligible.  In the microcanonical ensemble
(fixed total energy), this also means that the wave functional
$\Psi[\varphi(\xi),\tau]$ describing the quantum state in the
Schr\"odinger representation has a weak time-dependence \cite{shanki}.
Under canonical transformation and at fixed
Lema\^itre time Hamiltonian reduces to free scalar field
propagating in flat space-time \cite{sdssss}. 

\vskip 0.05in
\noindent
\underline{\bf Entanglement entropy computation --- procedure:}

First, we discretize the Hamiltonian on a spherical lattice with
number of points $N$ ($\gg 1$) and of spacing $a$ ($\ll L = (N+1) a$
--- the infrared cutoff), whence the Hamiltonian reduces to that of
$N$-coupled harmonic oscillators.

Then we choose, for simplicity, the quantum state described by $\Psi$,
to be in a superposition \cite{sdssss} of the vacuum (ground) state
and the 1-particle (excited) state
described by $\Psi_0$ and $\Psi_1$
respectively, i.e., $\Psi = c_0 \Psi_0 + c_1 \Psi_1$, with $|c_0|^2 +
|c_1|^2 = 1$.

Finally, we obtain the reduced density matrix by tracing over $n$ of
the $N$ oscillators\footnote{$n$ oscillators are supposed to provide
  the DOF inside the horizon $\rHo$, and the tracing is therefore over
  the region enclosing the horizon [$\xi \to (\rHo, \infty)$] in
  Lema\^itre coordinates \cite{sdssss}.}, and evaluate the entropy
$\SEN$ using the above Von Newmann relation. Due to extreme difficulty
in analytical computations, we adopt numerical techniques.

\vskip 0.05in
\noindent
\underline{\bf Entanglement entropy computation --- results:}

The best fit for the entanglement entropy for the
superposition (or mixing) of vacuum and 1-particle states gives a
power-law correction to $\SBH$ (see Fig. 1)\cite{sdssss}
{\small
\be
\SEN = S_{_{\rm BH}} \l[1 + a_1 \l(\frac{{\cA}_{\rm H}}
{\ell_{_{\rm Pl}}^2}\r)^{-\beta}\r] \qquad a_1, \beta > 0 \, .
\ee
}
The parameter $a_1$ increases with $|c_1|$; and when $c_1 = 0$,
$a_1 = 0$ and $\SEN = \SBH$. Thus, the entanglement entropy of the ground
state obeys the area law. Only the excited state contributes to the
correction, and more excitations produce more deviation from the area law.
The parameter $\beta$ lies between $1$ and $2$. So for large BHs (i.e.,
large ${\cA}_{\rm H}$) the correction term falls off rapidly and the area 
law is recovered, whereas for the small BHs the correction is significant.
\begin{figure}
\begin{center}
\epsfxsize 100mm
\epsfysize 38mm
\epsfbox{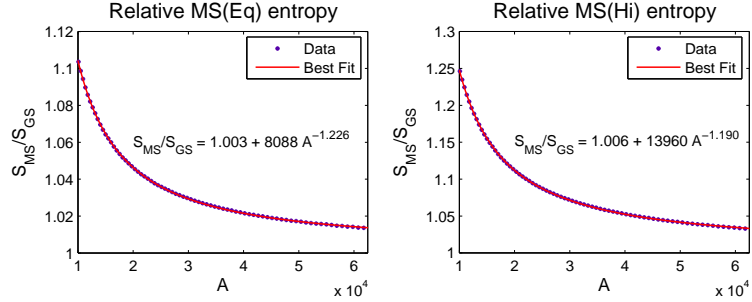}
\caption{\small Best fit plots (solid lines) of the 
relative mixed state entropies ($S_{_{\rm MS}}/S_{_{\rm GS}}$) for 
equal and high mixings versus the area $\cA$. The corresponding data are 
shown by asterisks.}
\la{fig:MS-fit}
\end{center}
\vspace*{-0.55cm}
\end{figure}
This can be interpreted as follows: for large ${\cA}_{\rm H}$, i.e.,
at low energies, it is difficult to excite the modes and hence, the
ground state modes contribute to most of the $\SEN$. However, for
small horizon area, a large number of field modes can be excited and
contribute significantly to the correction causing large deviation
from the area law. In fact, the increase in the deviation with
excitations may be attributed to the fact that the total entropy gets
increasing contributions from the quantum field DOF that are far from
the horizon, rather than immediately close to it\footnote{The
  near-horizon DOF of course contribute to the bulk of the entropy
  even for ${\cA}_{\rm H}/\ell_{_{\rm Pl}}^2 \sim {\cal O} (10)$, thus
  keeping the power-law correction sub-leading for fairly small-sized
  black-holes.}. See refs \citen{sdssdof,sdssssdof} for details.

In conclusion, let us make the following remarks: (i) As $\SEN$ is
obtained for a scalar field in flat space-time, $\SBH$ and its
correction can be identified uniquely with the quantum state
correlations. (ii) Although we have considered the microcanonical
ensemble, the identification of the power-law correction to the
kinematical properties of the horizon can be done by obtaining $\SEN$
in the canonical ensemble \cite{shanki,sarkar}.

SSh is supported by Marie Curie Incoming International Grant IIF-2006-039205.


\end{document}